\documentclass[graybox]{svmult}

\usepackage{helvet}         
\usepackage{courier}        
\usepackage{newtxtext,newtxmath} 
\usepackage{type1cm}        
\usepackage{makeidx}         
\usepackage{graphicx}        
\usepackage{multicol}        
\usepackage[bottom]{footmisc}
\usepackage[utf8]{inputenc}
\usepackage{float}
\usepackage{subfigure}  
\usepackage{multirow}
\usepackage{capt-of}
\usepackage{comment}

\usepackage[absolute]{textpos}
\makeindex             

\begin{document}

\title*{{VINEVI}: A {Virtualized} {Network} {Vision} {Architecture} for {Smart} {Monitoring} of {Heterogeneous} {Applications} and {Infrastructures}}


\author{Rodrigo Moreira, Hugo G. V. O. da Cunha, Larissa F. Rodrigues Moreira, Fl{\'a}vio de Oliveira Silva}

\institute{Rodrigo Moreira, \at Institute of Exact and Technological Sciences (IEP), Federal University of Viçosa, \\ Rio Paranaíba, Brazil, 38810-000, \email{rodrigo@ufv.br} \and Hugo G. V. O. da Cunha, \at Faculty of Computing (FACOM), Federal University of Uberl{\^a}ndia, \\  Uberl{\^a}ndia, Brazil, 38400-902,  \email{hugo.cunha@ufu.br} \and Larissa F. Rodrigues Moreira, \at Faculty of Computing (FACOM), Federal University of Uberl{\^a}ndia, \\  Uberl{\^a}ndia, Brazil, 38400-902,  \email{larissarodrigues@ufu.br}
\and Fl{\'a}vio de Oliveira Silva, \at Faculty of Computing (FACOM), Federal University of Uberl{\^a}ndia, \\ Uberl{\^a}ndia, Brazil, 38400-902, \email{flavio@ufu.br}}

\maketitle

\begin{textblock*}{15cm}(3cm,25cm) 
\noindent
    \textcolor{red}{This paper has been accepted by the 36th International Conference on Advanced Information Networking and Applications (AINA-2022). The definite version of this work was published by Springer as part of the AINA 2022 proceedings. 
    DOI: \url{https://doi.org/10.1007/978-3-030-99584-3\_46}.}
\end{textblock*}

\abstract{
Monitoring heterogeneous infrastructures and applications is essential to cope with user requirements properly, but it still lacks enhancements. The well-known state-of-the-art methods and tools do not support seamless monitoring of bare-metal, low-cost infrastructures, neither hosted nor virtualized services with fine-grained details. This work proposes VIrtualized NEtwork VIsion architecture (VINEVI), an intelligent method for seamless monitoring heterogeneous infrastructures and applications. The VINEVI architecture advances state of the art with a node-embedded traffic classification agent placing physical and virtualized infrastructures enabling real-time traffic classification. VINEVI combines this real-time traffic classification with well-known tools such as Prometheus and Victoria Metrics to monitor the entire stack from the hardware to the virtualized applications. Experimental results showcased that VINEVI architecture allowed seamless heterogeneous infrastructure monitoring with a higher level of detail beyond literature. Also, our node-embedded real-time Internet traffic classifier evolved with flexibility the methods with monitoring heterogeneous infrastructures seamlessly.
}

\section{Introduction}\label{sec:introduction}


Understanding how Internet services and resources are used is essential to support the user experience within Service-Level Agreement (SLA)~\cite{Manisha2018}. Among the Internet resources, the cloud computing infrastructures, which use virtualization and handle vast amounts of data generated by users, require the entire stack monitoring from the hardware to the virtualized applications. In 2014~\cite{etsi2014}, the cloud computing paradigm evolved to Multi-Acces Edge Computing (MEC) to address the challenges of having centralized computing capabilities geographically distant from users~\cite{Pawani2018}.



Monitoring these infrastructures, especially cloud computing, has become essential for maintaining the service's operation, yielding users' service level agreements~\cite{Nahantesh2020}. Also, monitoring is critical to support visibility regarding their resource consumption behavior, enabling the prediction of outages, perform performance diagnosis, and Service Level Agreement (SLA) violation~\cite{Alessandro2019, Martin2020, Maria2021}. Furthermore, due to the large amount of data that monitoring tools generate, it becomes challenging to find methods, frameworks, or tools that detail the status of infrastructure entities, especially bottlenecks, without causing a significant overhead on the system~\cite{Popiolek2021}. 


Besides, monitoring heterogeneous infrastructures  Methods, technologies, and monitoring strategies found in the literature are predominantly integrated with cloud provider tools or tied to a specific SLA~\cite{Dylan2008, Carlos2014, Priscila2021} and do not take into account container-native monitoring~\cite{Taherizadeh2018}.


Hence, this work proposes the \textbf{VI}rtualized \textbf{NE}twork \textbf{VI}sion architecture (VINEVI) framework for seamless monitoring of heterogeneous infrastructures and services. VINEVI provides a set of entities and technological enablers that allow monitoring cloud infrastructures such as heterogeneous bare-metal (x64) and low-cost (AArc64)~\cite{HugoCunha2021} architectures. VINEVI advances state of the art with a monitoring method based on Artificial Intelligence (AI) that monitors the infrastructures, considering their resources and services. VINEVI enables the monitoring of network traffic volume by application class for each monitored entity and service. Additionally, this paper innovates with a framework that monitors hosted or virtualized services.


Among the contributions of this work, the following stand out: 


\begin{itemize}
    \item A seamless monitoring framework for network entities, hybrid infrastructures, and hosted and virtualized services;
    \item A traffic volume counter customizable by application class for hybrid architectures;  
    \item An assessment of the performance of CNNs as enabling technologies for real-time sampling network traffic classification.
\end{itemize}

This work is organized as follows. Section~\ref{sec:related_work} presents the current state of the art on network monitoring. Section~\ref{sec:vinevi_arch} presents the VINEVI architecture proposed in this paper, whereas the experimental setup as described in detail in Section~\ref{sec:experimental_setup}. Section~\ref{sec:results_and_discussion} reports results and analysis of the experimental evaluation. Section~\ref{sec:conclusion} concludes the paper.


\section{Related Work}\label{sec:related_work}


Currently, the literature presents different monitoring solutions for virtualized networks, and infrastructures \cite{Kreutz2015} \cite{Tsai2018} \cite{Zhou2018}. Some approaches focus on networks, others on infrastructures, but none rely on seamless monitoring of low-cost and high-performance infrastructures. Also, the well-known monitoring solutions lack monitoring both running on top of virtualized infrastructures or non-virtualized. This section describes some related work considering the seamless infrastructures and network monitoring capabilities.



Borylo et al.~\cite{Borylo2021} proposed and evaluated a portable monitoring module that combines monitoring capabilities to Network Function Virtualization (NFV), Software-defined Networking (SDN), and Cloud. Their solution architecture receives monitoring statistics through well-known and universal interfaces with SDN controllers, Virtualized Infrastructure Manager (VIM), host, and tenants. Unlike our work, we proposed a seamless infrastructure monitoring that considers a virtualized and non-virtualized infrastructure and its services. Although, we provide fine-grained statistical monitoring of the entire network and infrastructure. 


In \cite{Mfula2021} contains the description of a seamless platform based on Prometheus and Grafana for the deployment and monitoring of containers over infrastructures. Unlike our proposal, the authors focused on monitoring containers to achieve availability and OPEX reduction. Although our monitoring proposal can monitor container-based services, we go further by intelligently monitoring network traffic. VINEVI monitors bare-metal, low-cost infrastructures, network entities, and services transparently, unlike previous approaches. Similarly, we find other monitoring architectures in \cite{Mfula2021} which take into account logs~\cite{Nitin2020} while~\cite{Ahmed2019} takes into account flow rules.


Won \& Kim~\cite{Hojoon2021} believe that open source tools like Prometheus and Zabbix require configuration and rely on empirical knowledge about failures, requiring administrators to set accurate thresholds for each situation. They proposed an intelligent Multi-Layer monitoring architecture based on Prometheus and machine learning to address these challenges. The authors evaluated the efficiency of some machine learning models such as Random Forest (RF), Support Vector Machine (SVM), and Deep Neural Network (DNN) for monitoring tasks. The results suggest that DNN proved to be a promising technology with reasonable accuracy for predicting CPU, RAM, and network failures. Our paper also deals with statistics gathering and proactive submission to related endpoints. However, the VINEVI framework thoroughly monitors the infrastructure, which can be bare-metal, low-cost, network entities, virtualized, or hosted services on top of those infrastructures.


In~\cite{Jerico2021} we find DynAMo, an alternative monitoring approach to Prometheus that causes low computational overhead on monitored entities and services. The use case used in the evaluation refers to the communication service integrated into a railway company simulated in a software environment. Among the results reported, the low memory consumption compared to monitoring based on Prometheus stands out. Unlike the authors, VINEVI, in addition to monitoring containers in production environments, monitors the consumption of network resources of containers.

\section{VINEVI Architecture}\label{sec:vinevi_arch}
 

This work proposes a method for seamlessly monitoring bare-metal, low-cost infrastructures and network elements. In addition to dealing with the infrastructure, our solution can deal with the services that run on these infrastructures, directly monitoring hosted, virtualized, and nested-virtualized services.



\begin{figure}[htb]
\begin{center}
\includegraphics[width=\textwidth]{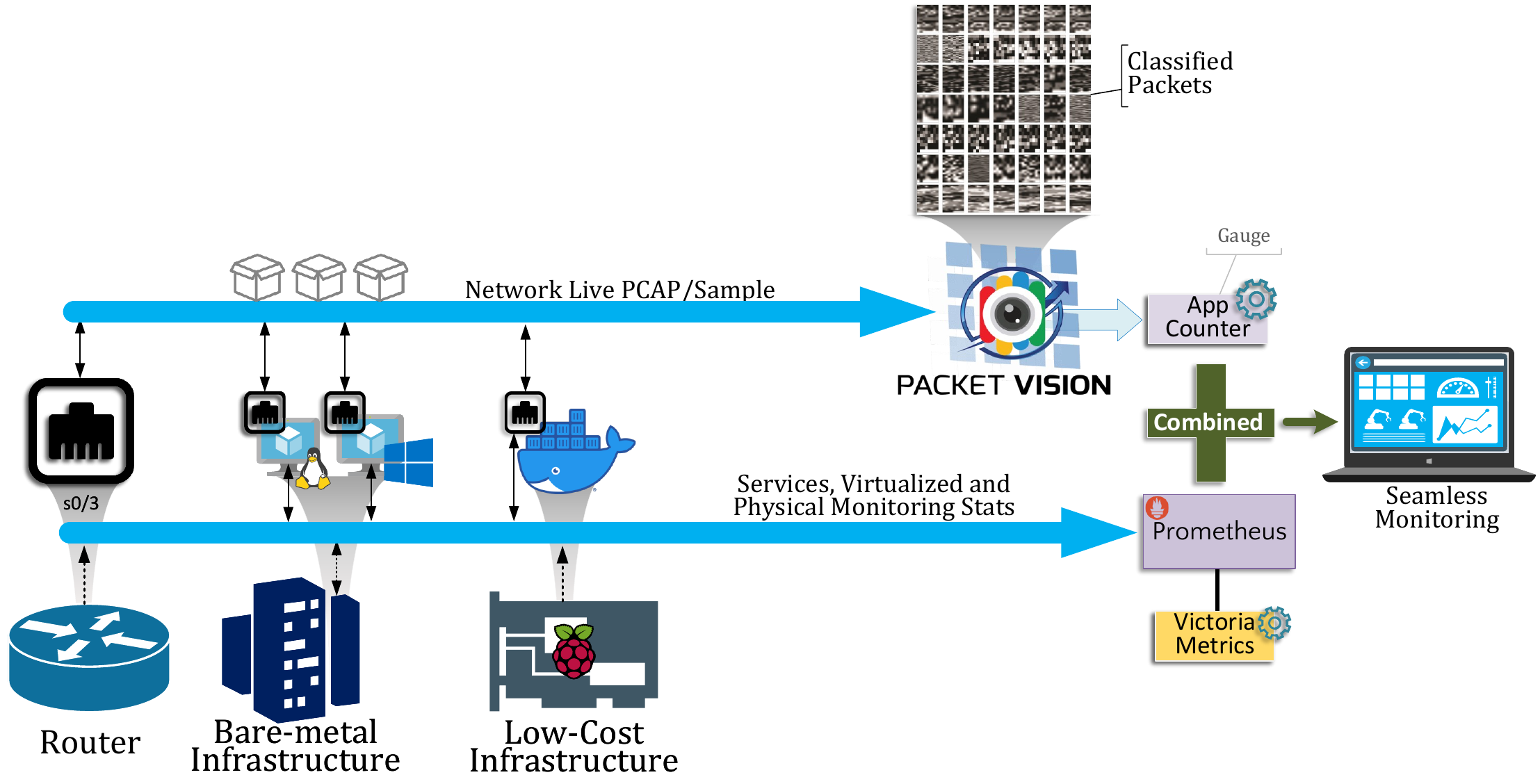}
\caption{VINEVI Monitoring Schema.}
\label{fig:vinevi_achitecture}
\end{center}
\end{figure}

VINEVI monitors the network elements and the traffic volume detailed by application class, services, and computing infrastructure. We present the conceptual diagram of the VINEVI framework in Fig.~\ref{fig:vinevi_achitecture} and read it from left to right, and we see network resources, data center infrastructure, and low-cost infrastructures.



In the upper flow, the blue arrow refers to the collection of information about network traffic of resources and services running on the infrastructures. In Fig.~\ref{fig:vinevi_achitecture} the \textit{s0/3} interface denotes the monitoring of the traffic volume of the router network interface, which can occur live or by samples. These samples feed the previously built~\cite{Moreira2020} Packet Vision component for traffic classification based on \textit{Convolution Neural Network (CNN)}. 

In the lower flow, information regarding the physical resources of the infrastructure feeds the event-driven statistics collector based on Prometheus~\cite{9060302}. In addition to physical information, the metrics collector receives information related to the service, depending on the service operator specifying what will be monitored from the application and pointing to the corresponding end-point. In addition to monitoring the infrastructure, VINEVI can monitor the service orchestrators of these infrastructures and the Virtualized Infrastructure Managers (VIMs).


The VINEVI framework combines service and resource monitoring metrics with specified traffic volume by application class. When combined, we bring novelty with a seamless monitoring method that allows administrators and the infrastructure operations team to monitor their resources adaptively and granularly.
 

A critical component of the VINEVI framework is PacketVision~\cite{Moreira2020}, which receives sampled or live stream network packets and classifies them according to their application class. In VINEVI, we build a gauge-type counter for each predominant Internet traffic class. This counter decreases or increases over time depending on the current state of the network. The traffic classes for our VINEVI Proof-of-Concept (POC) are detailed in Section~\ref{subsec:dataset_description}. Another fundamental component of VINEVI is Prometheus or Victoria Metrics which receive metrics by end-points and store them for temporal analysis. 


For the VINEVI framework to consolidate the various monitoring metrics, each monitored entity must have a statistics publisher that runs as a daemon. This monitoring entity sends the metrics to the corresponding endpoints in Prometheus.

\section{Experimental Setup}\label{sec:experimental_setup}

\begin{figure}[!htbp]
    \centering
    	\includegraphics[width=0.85\textwidth]{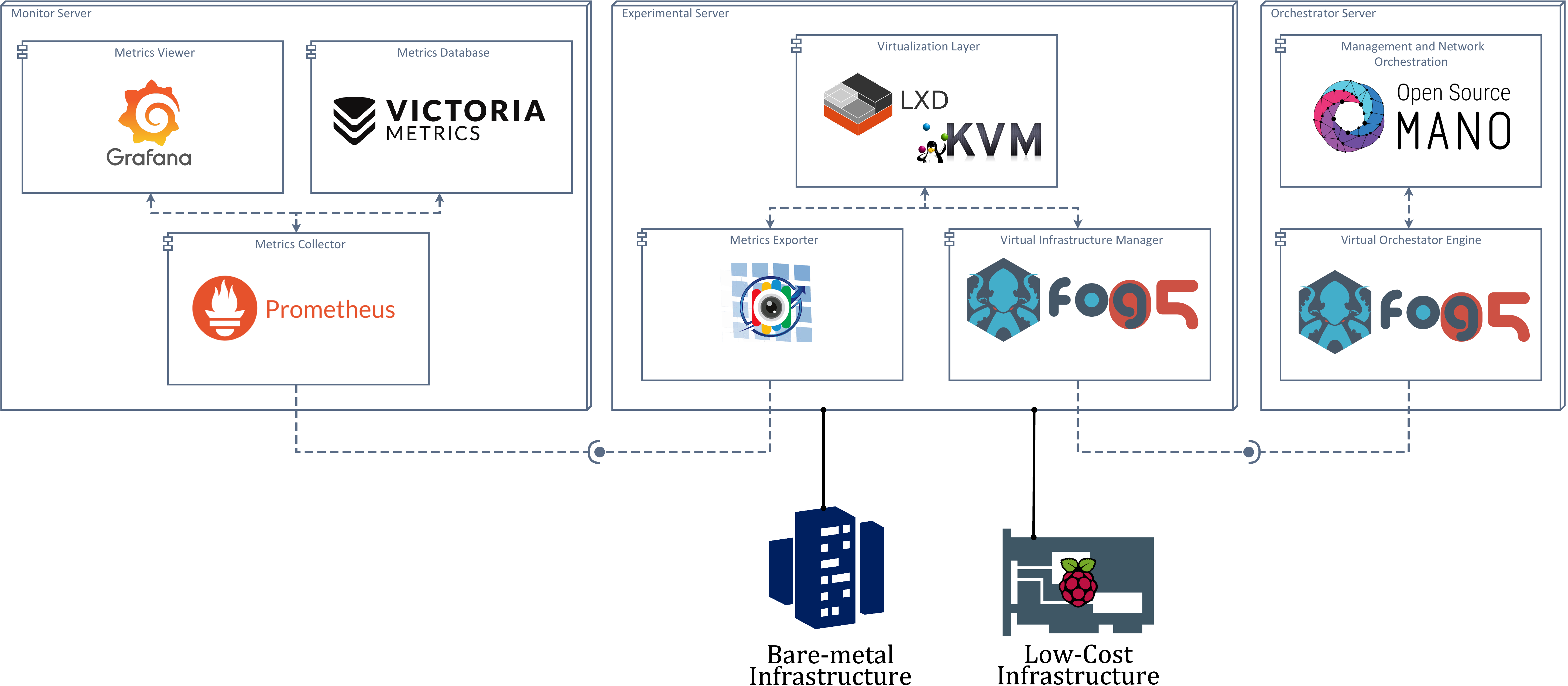}
    \caption{Intelligent Monitoring Testbed Overview.} 
    \label{fig:monitoring_schema}
\end{figure}

To functionally validate VINEVI, we deploy it in an experimental testbed. This testbed consisted of four (4) distinct hosts: the Monitor Server, Experimental Server, Orchestrator Server, and AI Server as in Fig.~\ref{fig:monitoring_schema}. Among these hosts represent the infrastructure in the VINEVI framework, the Experimental Server, a bare-metal with four vCPUs and 8 GB RAM with Ubuntu 18.04 LTS. Another type of Experimental Server admitted by VINEVI is the low-cost one. A Raspberry Pi4 is hosting virtual machines and is managed by Orchestrator Service.


The Monitor Server hosts the platform for viewing and supporting the monitoring data. We configure Grafana, Prometheus, and Victoria Metrics services to handle monitored entities. Besides, we configure a Virtual Private Network (VPN) service so that metric collectors publish metrics to endpoints correctly, even when they are outside the network domain. Part of our VINEVI experimental setup was deployed on the Microsoft Azure cloud computing platform with Ubuntu 18.04 LTS with a flavor 2vCPU and 8 GB RAM.



On the other hand, Orchestrator Server runs on a bare-metal server. Among the roles of this server, the implementation of virtualized services on bare-metal and low-cost infrastructure stands out. To seamlessly deploy virtual machines on bare-metal (x86\_x64) and low-cost (AArc64) infrastructures, we configured the open-source Eclipse fog05 virtual Orchestrator. This Orchestrator deals directly with the Fog5 agent running on top of Experimental Servers enabling them to launch virtual machines or containers.

\subsection{Smart Traffic Monitoring}\label{subsec:traffic_monitoring}


For VINEVI to monitor the volume of network traffic by application class, we incorporated the capabilities of CNNs into the framework through Packet Vision. Thus, VINEVI has an intelligent node agent for monitoring network traffic. We train CNNs to classify network traffic into seven (7) typical Internet classes: Bittorrent, Browsing, DNS, IoT, RDP, SSH, and VoIP. In this experimental setup we considered three (3) CNNs to classify traffic on the VINEVI framework: SqueezeNet~\cite{Iandola2016}, ResNet-18~\cite{He2015} and MobileNetV2~\cite{Sandler2018}. 



Among these three CNNs, the SqueezeNet and MobileNetV2 architectures were explicitly designed for use in mobile and edge devices, so we hypothesized that they are good candidates to compose the VINEVI monitoring agent. SqueezeNet aimed to reduce the number of parameters through fire modules, which use the strategy of compression and expansion of activation maps in the convolution layers~\cite{Iandola2016}. MobileNetV2~\cite{Sandler2018} uses separable depth convolutions, which consists of factoring the standard convolution into a depth convolution followed by a 1 $\times$1 convolution, called a point convolution.


The literature claims that it is not promising to embed large models such as AlexNet or VGG-16 in small devices because they demand high computational load \cite{Alippi2018}. However, we decided to investigate how the ResNet-18 architecture, which is the smallest network in the ResNet family (composed of $\approx 11$ million parameters), behaves when embedded. ResNet is composed of residual blocks that allow accelerating convergence and better deal with the \cite{He2015} overfitting problem.


A premise of the VINEVI framework is a previous CNN training to enable traffic prediction by application class for the monitoring agent. Once the model has been trained, the CNN has uploaded to the VINEVI intelligent traffic monitoring agent for future on-the-fly predictions on the infrastructures. The Torch framework allows loading previously trained models, avoiding training bottlenecks. Furthermore, due to the hardware restrictions of low-cost infrastructures, embedding the previous trained CNN model proves to be advantageous and functionally correct.


We adopted this strategy to enable the smart traffic monitoring agent on RPi4 to enrich infrastructure monitoring with application class details on traffic volume. We configured PyTorch 1.10.0 to run on AArc64 and x64 hardware, thus enabling the loading of an already trained CNN and for the monitoring agent to predict the sample packages' application class regardless of the architecture or underlying architecture infrastructure.


\subsection{Dataset Description}\label{subsec:dataset_description}


The dataset images evaluated in this paper were built from the Packet Vision~\cite{Moreira2020} component, which considered the raw information carried in the packet, including header and payload. About 9645 images were obtained from Wireshark traces (\textit{pcap}) from different sources. These images are categorized into seven (7) classes, as summarized in Table~\ref{tab:dataset}. Fig.~\ref{fig:dataset} shows some images from the dataset for each class. All images are in PNG format with $224 \times 224$ pixels size.

\begin{minipage}{1.0\textwidth}
 \begin{minipage}[c]{0.45\textwidth}
 \captionof{table}{Descriptive Summary of the images.}
 \label{tab:dataset}
 \resizebox{\textwidth}{!}{
 \begin{tabular}{lcl}
\hline
\multicolumn{1}{c}{\textbf{Class}} & \textbf{Samples} & \multicolumn{1}{c}{\textbf{Source}} \\ \hline
Bit Torrent & 1217 & \multirow{2}{*}{UPC Data \cite{Valentin2014}} \\
DNS & 1412 &  \\ \hline
Browsing & 1225 & \multirow{3}{*}{ISCXVPN2016 \cite{Draper-Gil2016}} \\
RDP & 1271 &  \\
SSH & 1352 &  \\ \hline
IoT & 1848 & IoT Sentinel \cite{Miettinen2017} \\ \hline
VoIP & 1320 & NASOR \cite{Moreira2021} \\ \hline
\textbf{Total} & \multicolumn{1}{c}{\textbf{9645}} \\ \hline
\end{tabular}}
 
 \end{minipage}
 \hfill
 \begin{minipage}[c]{0.49\textwidth}
 \centering
 \centering
 \includegraphics[width=0.8\textwidth]{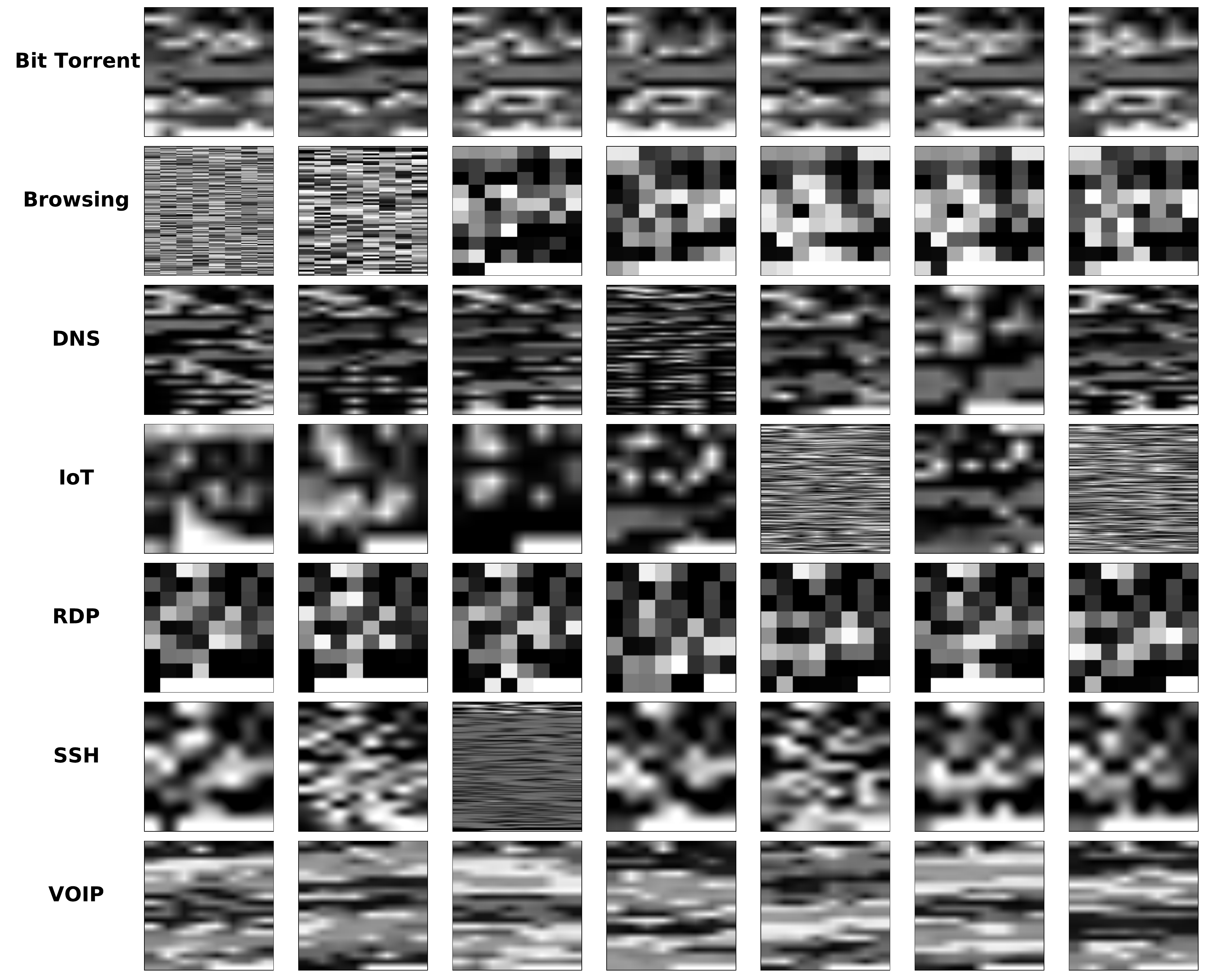}
 \captionof{figure}{Examples of images for each class.}
 \label{fig:dataset}
 \end{minipage}
\end{minipage}

\section{Results and Discussion}\label{sec:results_and_discussion}

Considering the testbed presented above, we carried experiments to validate the VINEVI framework functionally. The objectives of the experiments were to answer the following questions: 

\begin{enumerate}
    
    \item For the VINEVI testbed, which CNN outperforms regarding accuracy for real-time network traffic prediction by application class?
    
    
    \item For a real-time traffic prediction environment, where the prediction time of packets is essential, which CNN is suitable to compose the intelligent monitoring agent in AArc64 architectures?
    
    \item When does VINEVI intelligent monitoring agent run on x64 architectures, which CNN is best suited for traffic prediction by application classes?
    
    \item Is there differentiation in CPU consumption in the traffic class prediction process depending on CNN type and traffic class?
    
\end{enumerate}

\begin{figure}
\centering
\hfill
\subfigure[SqueezeNet]{\includegraphics[width=0.45\textwidth]{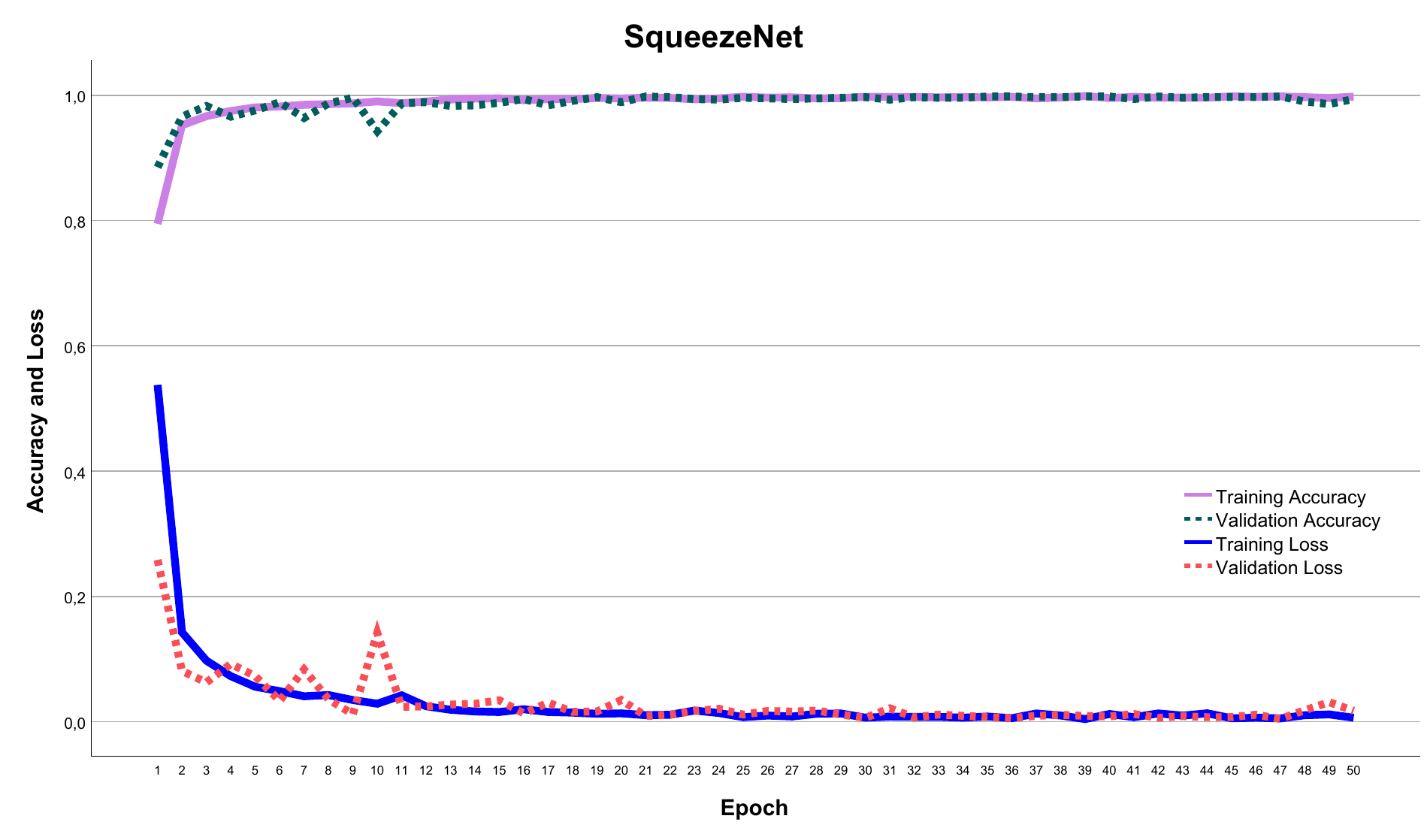}}
\hfill
\subfigure[MobileNet]{\includegraphics[width=0.45\textwidth]{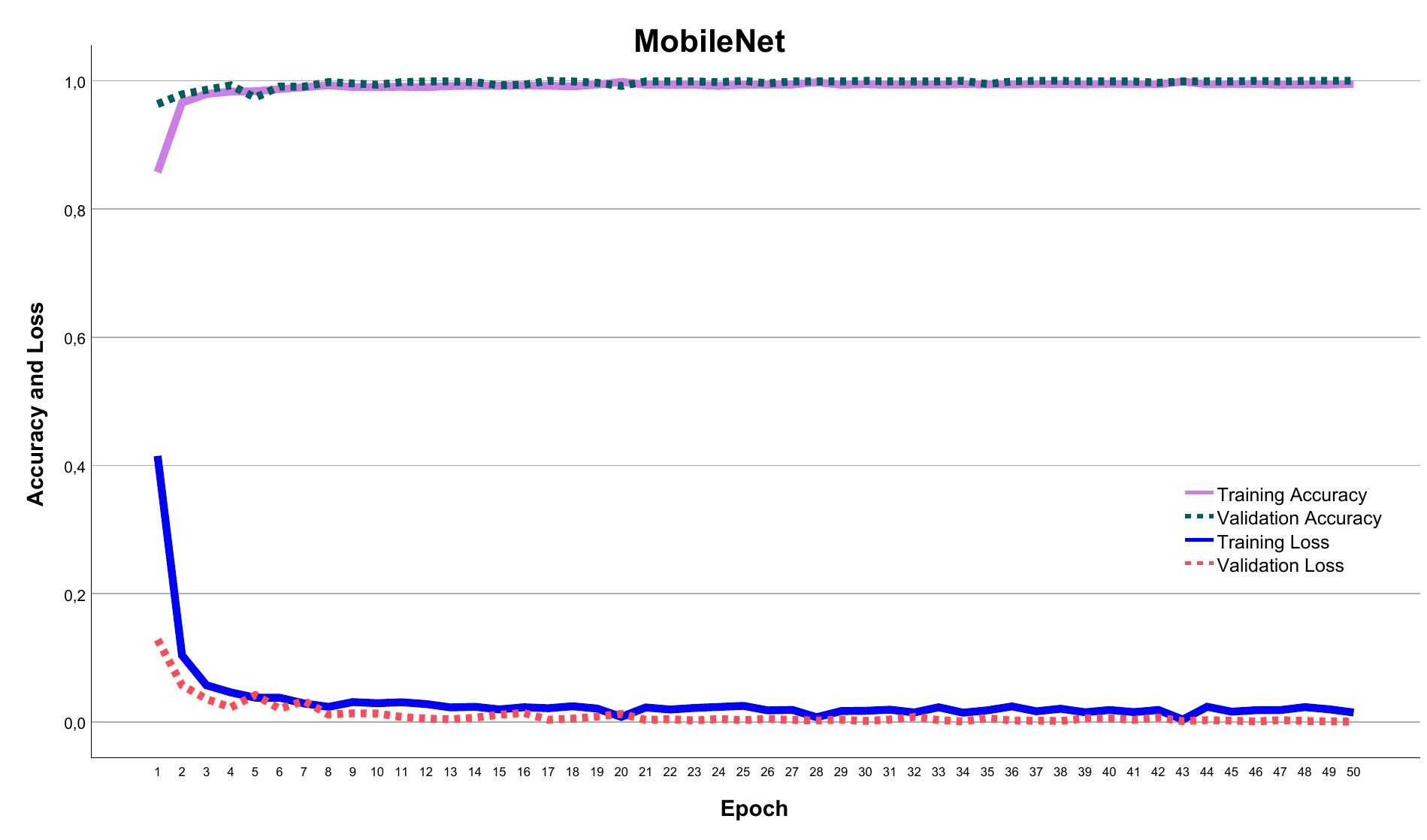}}
\hfill
\subfigure[ResNet]{\includegraphics[width=0.45\textwidth]{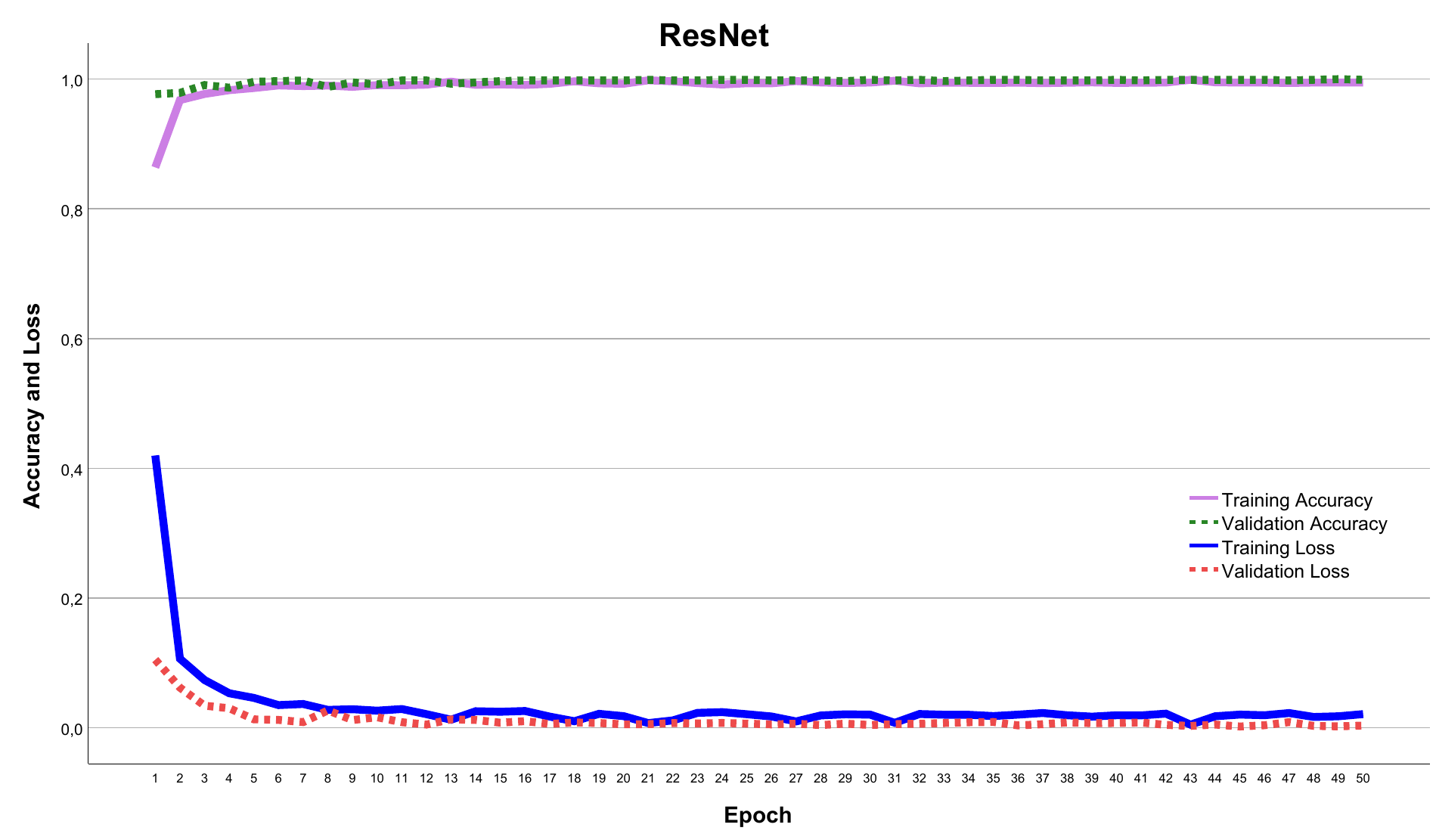}}
\caption{Training and Loss values evolution for each CNN model.}
\label{fig:chart_training}
\end{figure}

For VINEVI's experimental setup, and considering the dataset described above, the CNN that performed best regarding accuracy was MobileNet. The numerical results of the learning and testing process denoted that the accuracy of this CNN was $99.90\%$. The learning and prediction behavior, according to Fig.~\ref{fig:chart_training}, implies the generalization and aptitude of the model to compose the VINEVI smart traffic monitoring agent.


When we implemented the intelligent traffic classification agent running on low-cost and bare-metal infrastructures, we noticed a different behavior for the same CNN. Intelligently monitoring the traffic class of applications on low-cost infrastructures (AArc64) can be done by pooling sampling, not in real-time because of the large amount of data in a production network. To predict a single packet regarding its application class on AArc64 architectures required $\approx811ms$ using ResNet. Thus, ResNet was the CNN that required the shortest time to predict the application class of a given package in AArc64.


When the intelligent traffic prediction agent ran on bare-metal (x64) infrastructures, ResNet demanded the longest prediction time $\approx 77 ms$, behaving differently than AArc64 architecture. In the x64 architecture, the CNN that consumed the least amount of time to predict the application class of a given package was SqueezeNet, demanding around $\approx 62 ms$, according to Fig.~\ref{fig:x64}. In Fig.~\ref{fig:AArch64} it can be seen that CNN ResNet took about $9.5 \times$ less time than SqueezeNet to predict the traffic class of the same sampled network packet.


Accordingly, for VINEVI's testbed, the CNN outperforms prediction time depending on the architecture where the intelligent network traffic monitoring agent stands out. If deployed over low-cost infrastructures, the best CNN to be used will be ResNet; if deployed on bare-metal, the most recommended is SqueezeNet.

\begin{minipage}{0.9\textwidth}
\begin{minipage}[c]{0.49\textwidth}
 \centering
 \includegraphics[height=7.3cm,width=10cm,angle=0, trim=0cm 0.5cm 2cm 0cm,clip, width=1.0\textwidth]{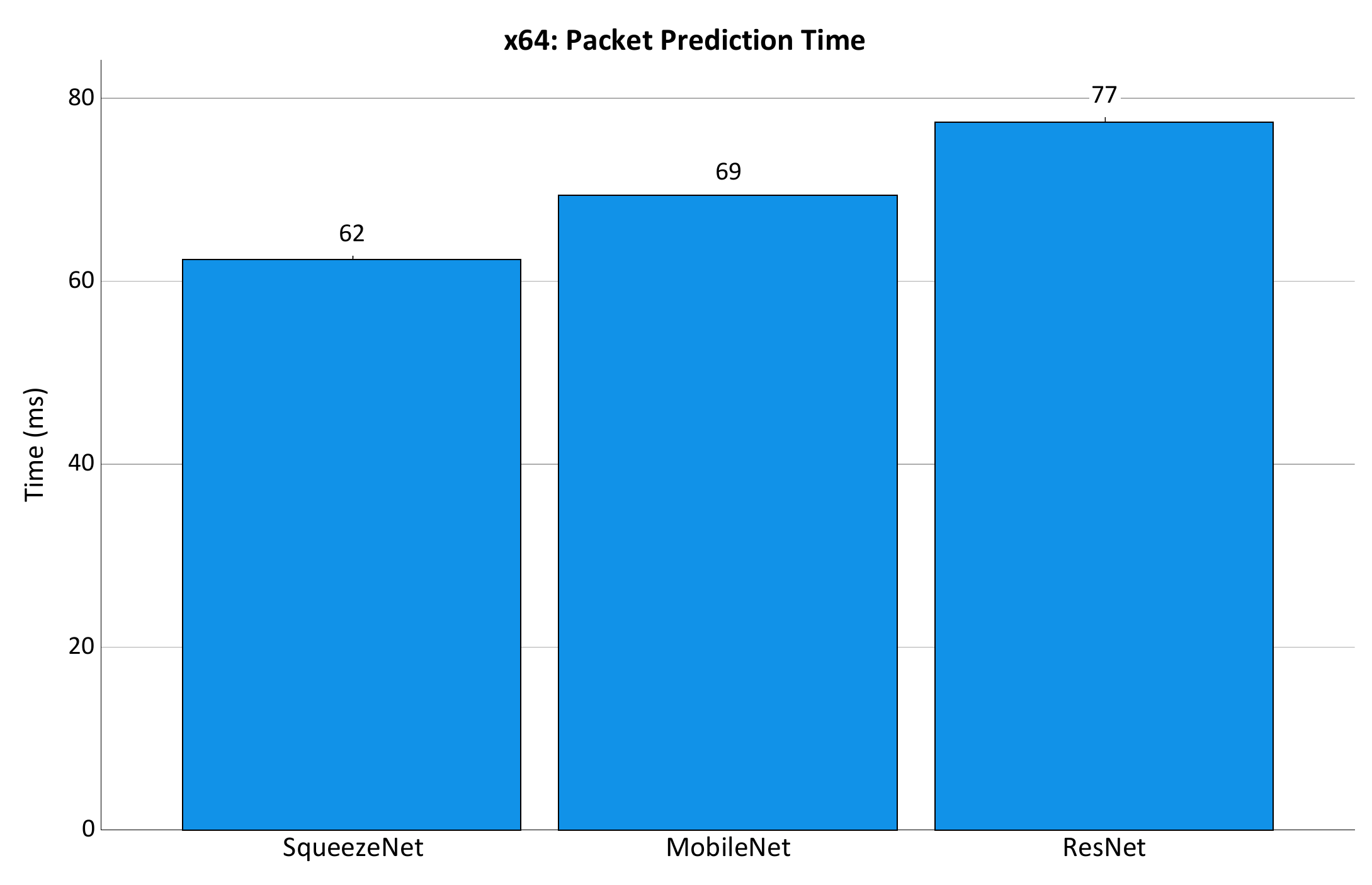}
 \captionof{figure}{x64: prediction time.}
 \label{fig:x64}
 \end{minipage}
 \hfill
 \begin{minipage}[c]{0.51\textwidth}
 \centering
 \includegraphics[height=7.3cm,width=10cm,angle=0, trim=0cm 4.2cm 2.5cm 0cm,clip, width=1.0\textwidth]{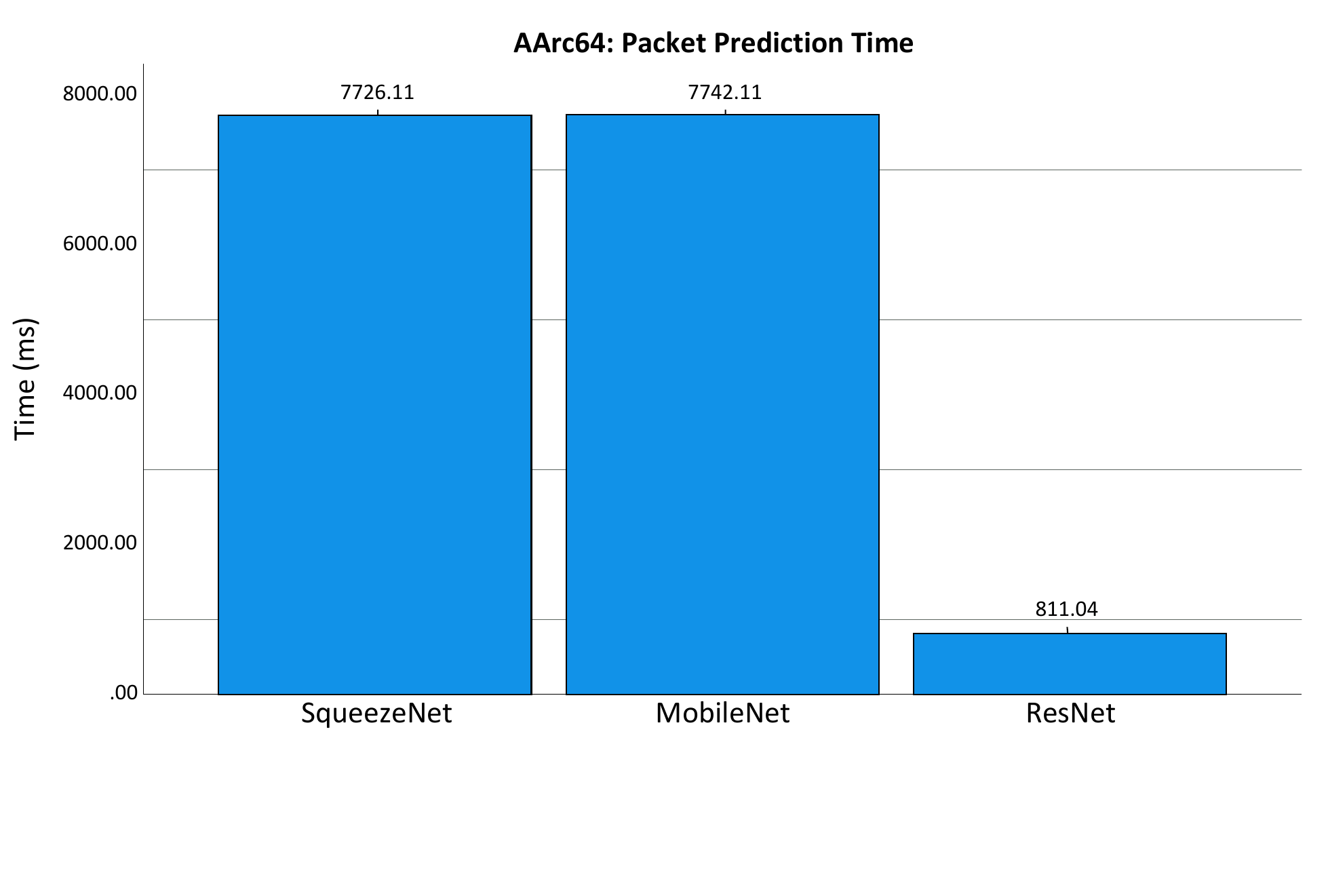}
 \captionof{figure}{AArch64: prediction time.}
  \label{fig:AArch64}
 \end{minipage}
\end{minipage}

\bigskip

\begin{figure}[!htb]
\begin{center}
\includegraphics[width=\textwidth]{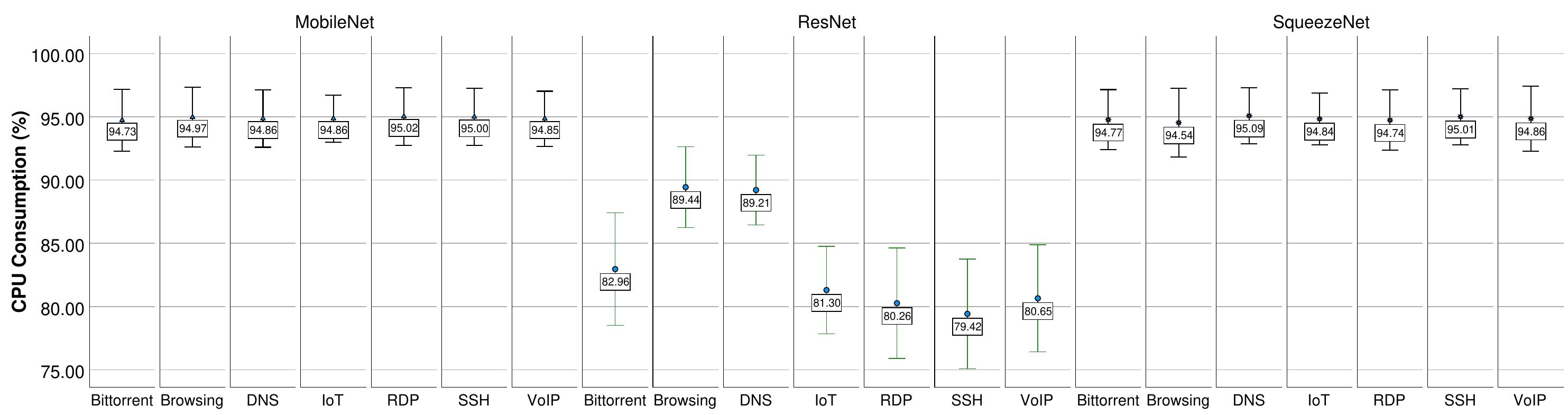}
\caption{Comparison of CPU consumption by classification agents.}
\label{fig:cpu_consumption}
\end{center}
\end{figure}

We investigate how the placement of intelligent traffic monitoring agents over low-cost infrastructure impacts the consumption of computational resources. In this investigation, we consider how and to what extent the prediction of different classes of applications by the three (3) CNNs considered in the VINEVI testbed affects the CPU consumption of low-cost infrastructures. According to Fig.~\ref{fig:cpu_consumption} it is possible to see that MobileNet and SqueezeNet consumed on average and within the confidence interval the same amount of CPU for all application classes.

However, with ResNet, it was observed that the average CPU consumption was $\approx 82.49\%$, being $\approx13\%$ lower compared to other CNNs. We believe that this happened because in the process and prediction, having the CNNs already been trained, the last SoftMax layer is responsible for the forecast. Since this layer is less complex, it implies lower CPU consumption. Therefore, incorporating this CNN in the intelligent traffic monitoring agent to monitor low-cost infrastructures is proved to be more suitable. 



We also assess the complexity of each CNN, and according to~\cite{Zhang2018} the complexity of a CNN is measured by the amount of multiplication and addition operations that are required for the underlying architecture to perform a computation~\cite{Ma2018}. According to Table~\ref{tab:complexity_cnn}, the last layer of ResNet is less complex in terms of FLOPS compared with others in the same experiment. Due to this lesser complexity, we argue that the ResNet consumed less CPU than others. Also, this result suggests that embedded systems can provide better performance during feedforward prediction~\cite{Dundar2017}.

\begin{table}[htbp]
\begin{center}
\caption{Complexity of the last layer of CNNs.}
\label{tab:computational_cost}
\begin{tabular}{lcc}
\hline
\multicolumn{1}{c}{\textbf{CNN}} & \textbf{No. of Parameters} & \textbf{Complexity (\%)} \\ \hline
SqueezeNet & 0.004 M & 0.082 \\
MobileNet & 0.009 M & 0.003 \\
ResNet & 0.004 M & 0.000 \\ \hline
\end{tabular}%
\label{tab:complexity_cnn}
\end{center}
\end{table}


The VINEVI framework showed to be adaptable to monitor network traffic and heterogeneous infrastructure resources seamlessly, not impeding various infrastructures that exist on the Internet. Thus, when we combine the traffic class predictor with well-established state-of-the-art monitoring mechanisms, we arrive at an AI-based solution that seamlessly monitors heterogeneous infrastructure and services.

\section{Conclusion}\label{sec:conclusion}


This work introduced the VINEVI framework for seamless monitoring of network traffic, hybrid infrastructures, and hosted or virtualized services. By combining the monitoring capabilities of well-established state-of-the-art tools~\cite{HugoCunha2021} with artificial intelligence technologies, we enrich state-of-the-art with detailed monitoring of hybrid entities and services. 



We proposed and functionally evaluated a counter gauge for monitoring network traffic volume by class of applications. This counter relies on CNN-based network traffic classification. In addition, we assessed the placement of the network traffic prediction agent on top of possible VINEVI monitoring architectures. We found that the traffic monitoring module must consider the underlying architecture type to load the CNN model, which takes the least time and consumes the least CPU to predict the application class of a given packet. Furthermore, seamless monitoring of infrastructures or services requires flexible solutions that are adaptable to the environment regardless of vendor, hardware, or software.



For future work, urges to try and validate other AI techniques for traffic prediction. In addition, we consider it essential to create standardized interfaces for infrastructure operation and automation solutions to use metrics monitored by VINEVI to positively impact metrics such as Mean Time to Failure (MTTF) or Mean Time to Recovery (MTR). Furthermore, we consider it necessary to study intelligent pooling and network sampling mechanisms for estimating traffic volume, class of applications, and others.

\begin{acknowledgement}
This study was financed in part by the Coordenação de Aperfeiçoamento de Pessoal de Nível Superior - Brasil (CAPES) - Finance Code 001. And we would like to thank National Education and Research Network (RNP) for financial support under the CT-Mon call.
\end{acknowledgement}

\bibliographystyle{ieeetr}
\bibliography{references}

\end{document}